**Fabrication of ultra thin single crystal diamond membranes \*\***

By *Barbara A. Fairchild\*, Paolo Olivero, Sergey Rubanov, Andrew D. Greentree, Felix Waldermann, Robert A. Taylor, Ian Walmsley, Jason Smith, Shane Huntington, Brant Gibson, David N. Jamieson,* and *Steven Prawer*

[\*]     Ms B.A. Fairchild Corresponding-Author, Dr. P. Olivero†, Dr S. Rubanov, Dr A.D. Greentree, Prof. D.N. Jamieson, Prof. S. Prawer
School of Physics,
University of Melbourne, Melbourne, 3010 (Australia)
E-mail: babs@physics.unimelb.edu.au

       Mr F. Waldermann, Dr R. A. Taylor, Prof. I. Walmsley,
Clarendon Laboratory, University of Oxford, Oxford, OX1 3PU (United Kingdom)

       Dr J. Smith
Department of Materials
University of Oxford, Oxford, OX1 3PH (United Kingdom)

       Dr S. Huntington, Dr B. Gibson
Quantum Communications Victoria,
School of Physics,
University of Melbourne, Melbourne, 3010 (Australia)

[\*\*]   This work was supported by the Australian Research Council, the Department of Education, Science and Training, the Australian Government, International Science Linkages program and by the US National Security Agency (NSA), and the Army Research Office (ARO) under Contract No. W911NF-04-1-0290. ADG is the recipient of an Australian Research Council Queen Elizabeth II Fellowship (project number DP0880466). QCV acknowledges support from the Victorian Government's Science, Technology and Innovation Infrastructure Grants Program.

The extreme properties of diamond have led its use in a wide variety of applications from drilling for oil to ultra sharp knives for eye surgery. In the nanotechnology realm, diamond is being recognized as a material with great potential. For example, the extremely high Young's modulus suggests that diamond could be used in Nano-Electro-Mechanical systems (NEMS) and quantum NEMS applications since diamond cantilevers will have higher oscillation frequencies than other materials. Furthermore diamond is chemically inert and biocompatible, and the surface can be functionalized which makes it attractive for potential nano-bio

---

† now at Experimental Physics Department, University of Torino, Torino, 10126 (Italy)

applications. There are many large dipole moment colour-centres in the visible range which can act as artificial atoms for quantum optics. In particular, diamond containing optically active colour centres (especially the negatively charged Nitrogen-Vacancy $NV^-$ and NE8 centres) is rapidly emerging as an ideal candidate for applications in quantum information processing (QIP).[1, 2] This promise is based on the proven room-temperature single spin coherence and readout properties of the ground states,[3] and optical spin polarization.[4] The optical centres are also extremely bright and photostable, leading to applications as single photon sources[5-8] for quantum key distribution and metrology applications. There is thus a strong motivation for the development of new techniques to process diamond into complex functional structures. The present work reports significant advances in realising such structures using ion beam techniques, demonstrating the formation of buried single crystal diamond membranes, suited for post-processing and liftout, with thickness down to 210 nm.

To optimize the collection of photons from individual colour centres, embedding the centres in cavity structures (preferably monolithic cavities) offers the possibility of improved collection efficiency and greater control of the optical transitions via the Purcell effect.[9-11] If the strong coupling regime can be reached, then quantum cavities and photonic interconnects can be realized, which would both be substantial enablers of a diamond-based quantum computer[12-14] or photonic condensed matter analog.[15-17] To date, whispering gallery mode resonators,[18] and photonic crystal microcavities[19] have been fabricated in nanocrystalline diamond thin films, and the coupling of nanodiamond to silica microspheres has also been reported.[20]

Against such promise are the extreme challenges raised by working with diamond. Until recently, materials of consistent quality were not available, and even selected natural diamonds did not have the levels of perfection required to demonstrate many desired features.

Diamond fabrication is relatively primitive compared with silicon technology, and certainly multi-layer heterostructures have not been possible due to the lack of dopants that provide etchable layers. Free standing poly-crystalline diamond layers have been fabricated as thin as 80 nm[21] and numerous devices have been produced: photonic crystals,[22] cavities,[18] beam[23] and tuning fork resonators,[24] Micro-Electro-Mechanical Systems (MEMS) devices[25] and other novel structures.

However, especially for applications in quantum information, it is highly desirable to be able to fabricate structures and devices from the highest quality single crystal diamond. Here we present methods for the fabrication of thin (~200 nm) diamond membranes sculpted from single crystal diamond substrates using a novel two-energy ion implant process. This process is essential for reaching the thickness required to make photonic bandgap structures in membranes from the single crystal starting material. This imperative can be understood thus: for optimum cavity performance, the thickness of the membrane should be of the order of $\lambda/n$, where $n = 2.4$ is the refractive index.[26] For diamond NV- centres emitting at 637 nm, this translates into a thickness of order 200 nm. Such thin membranes are achievable using nanodiamond thin films deposited on a sacrificial layer of $SiO_2$. However, nanodiamond films at present do not possess the optical and electronic properties required to meet the demands for scalable quantum information applications, which includes the need for low-noise, reproducible environments for the colour centres (especially on the optical transitions), low strain and low optical scattering. Thus methods need to be developed to produce thin diamond membranes from high quality single crystal material. Our robust method demonstrates the fabrication of single crystal diamond membranes only 210 nm thick with lateral dimensions of the order of up to 0.3 x 0.3 mm. The structures are suitable for further processing, and we have milled rings, disks, cantilevers and other cavity structures into these thin membranes.

Parikh[27] developed a method for removing entire sheets of single crystal diamond from bulk samples based on MeV ion implantation followed by thermal annealing. It is well-known that MeV ions implanted into a substrate produce a high damage region at the end of range and that above a certain damage threshold (the critical dose, Dc) diamond[28] will convert to an etchable carbon layer (ECL) upon thermal annealing, as seen in **Fig. 1**. Dc is $1 \times 10^{22}$ vac/cm$^3$ for surface implantations, but Dc is observed to increase with increasing implantation depth.[29] This can be used as an effective method to create a sacrificial layer at a controlled depth from the surface. Multi-mode waveguides and prototypical cantilever structures have been fabricated using MeV ion implantation with Focused Ion Beam (FIB) patterning [30] and Reactive Ion Etching (RIE).[31]

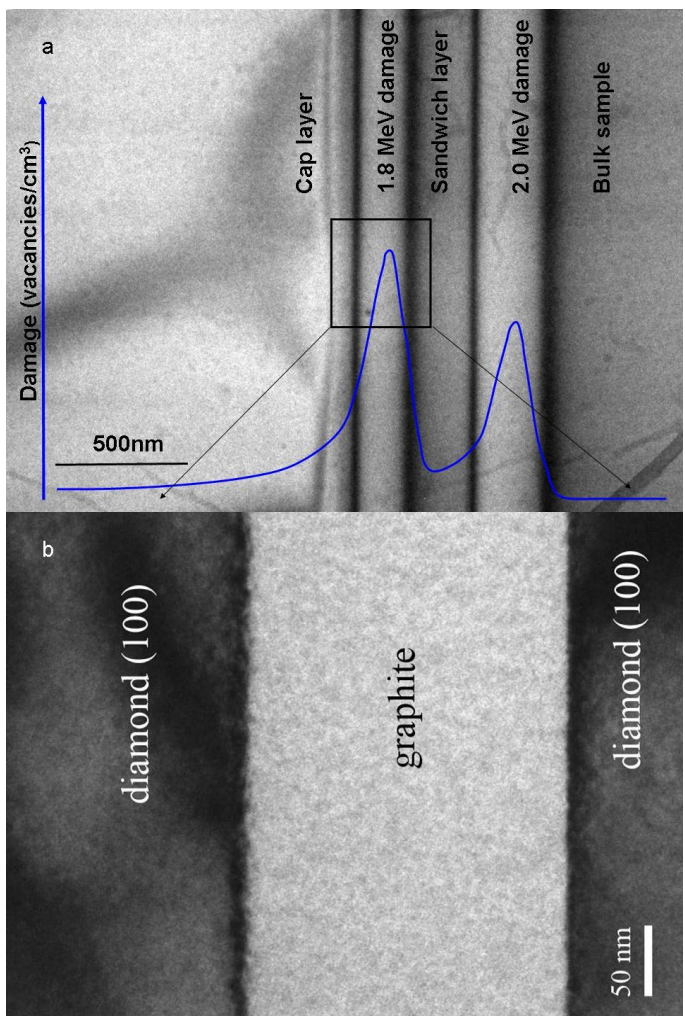

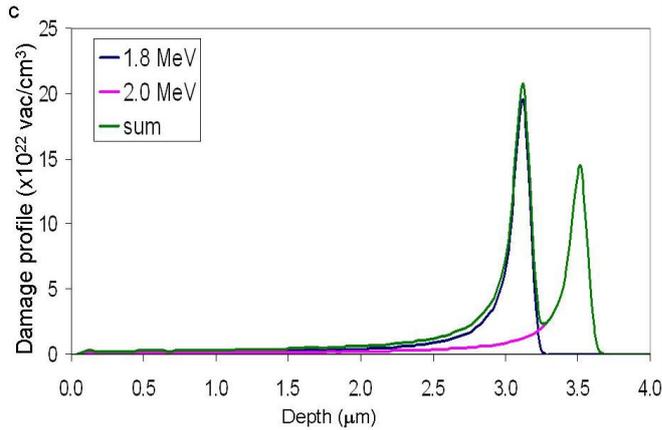

(**Fig. 1a**). Transmission electron microscopy (TEM) image of an unannealed, unetched implant with single crystal diamond and amorphous carbon zones. The sample is implanted from the left in this image. The cap layer and damage zones from the 1.8 and 2.0 MeV helium ions are marked. The 2.0 MeV ions are implanted first with the lower energy ions implanted afterwards. Note the darkened regions between the low damage and high damage regions, TEM indicates this is due to point defects in the diamond lattice.

(**Fig. 1b**). After annealing at 1260C the interface between the damaged and undamaged areas becomes significantly sharper. The highly damaged region converts to graphite and the transition zone between the graphite and diamond appears to be ≤ 10 nm.

(**Fig. 1c**). SRIM simulation of damage produced from two distinct ion implantations of helium ions into diamond. The 2.0 MeV helium ions, with fluence of $5 \times 10^{16}$ ions/cm$^2$ (pink curve) deposit their energy at greater depth (~3.5 μm) than the 1.8 MeV helium ions, with fluence of $6.8 \times 10^{16}$ ions/cm$^2$ (blue curve) at a shallower depth (~3.1 μm). The cumulative damage (green curve) is also shown.

The methods described in[30] are limited, however, in that it is not possible to reduce the thickness of the preserved cap layer to thicknesses compatible with single-mode visible waveguide operation, i.e. of order 200 nm. This is due to a combination of the longitudinal straggle and difficulties in controlling the strain within the layer, leading to cracking. One approach to overcoming these difficulties is the growth of a layer of CVD diamond on top of the implanted region.[32] Our approach is to employ a double implant strategy to create two damage layers with a single-crystal layer 'sandwiched' between them. SRIM simulations of the profiles for the expected vacancy distribution from a double implantation are shown in **Fig. 1c** where the implant species were 2.0 and 1.8 MeV Helium ions. The 'sandwich' layer lies between the maxima in the damage profile where the damage concentration is less than Dc. Fig. 1a shows a cross sectional TEM image of a sample prior to annealing and etching. The single crystal layer, approximately 330 nm thick is sandwiched between two heavily damaged layers in which the damage density exceeds Dc. The interface between the single crystal regions (D <Dc) and the heavily damaged regions (D>Dc) is not sharp and consists of crystalline material rich in point defects which appear as dark bands in Fig. 1a. After annealing the interface sharpens considerably as the regions in which D < Dc are annealed back to diamond and the regions D > Dc are converted into an etchable $sp^2$ bonded form of carbon. The interface region, of order 10 nm thick, has dislocation loops present. **Fig. 1b** shows a cross sectional TEM of one of the interfaces after annealing at 1260 C. Note that after annealing the interface roughness appears to be less than a maximum of 10 nm.

Galvanic etching can then be performed to remove the graphite-like layers creating a diamond/air/diamond/air/diamond structure as is shown in **Fig. 2a**. High Resolution TEM (HRTEM) **Fig. 2b** and diffraction patterns **Fig. 2c** of the sandwiched membrane show the expected pattern from cubic diamond.

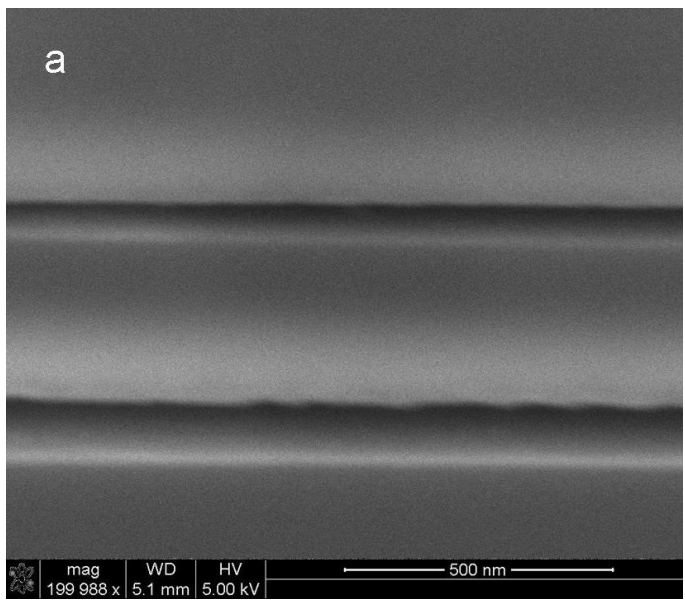

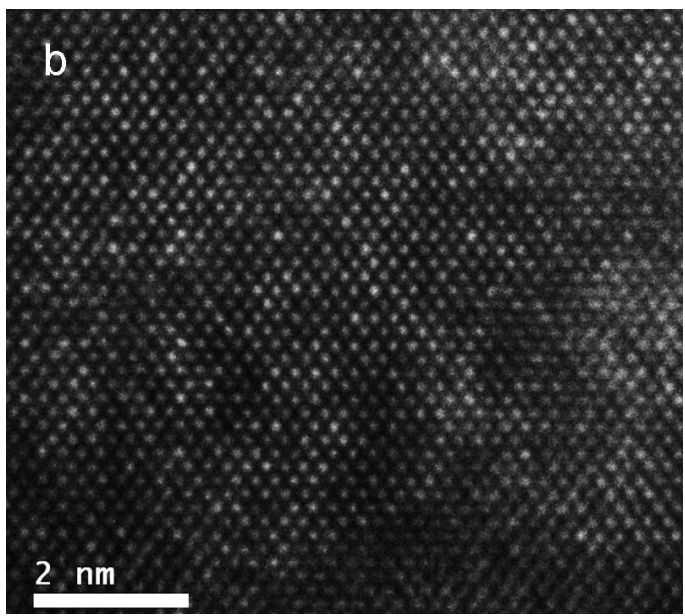

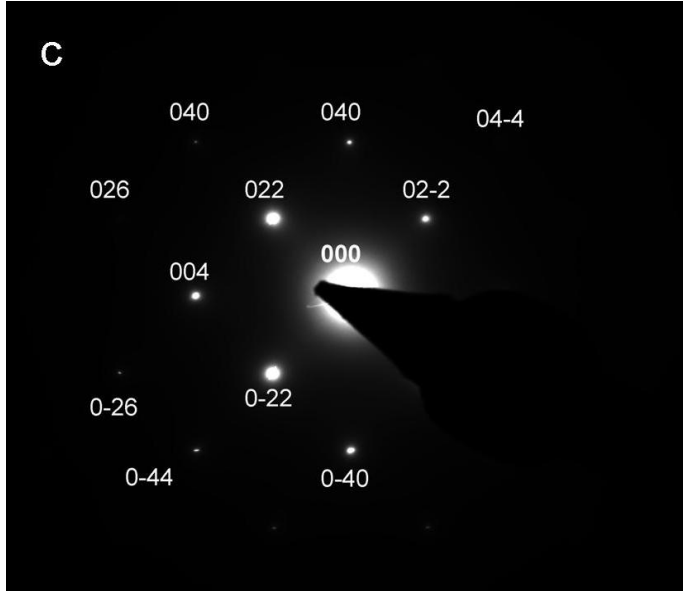

(**Fig. 2**). a) SEM image of 330 nm layer after etching showing ~100 nm air gaps between layers. b) A high resolution TEM (HRTEM) image of diamond layer produced using our method as implanted, before annealing and etching. (c) Diffraction image of diamond membrane after liftout with all spots indexed to single crystal diamond.

The double implant method can be used as the basis for the fabrication of a range of photonic and NEMS structures. A schematic of the fabrication process is shown in **Fig. 3**. Following the double implant and high temperature annealing and galvanic etching (Step 1), a focused ion beam (FIB) is used to carefully mill down to the depth of the shallower implant (Step 2). A sharp tip is then glued to the cap layer (Step 3) which is then lifted out (Step 4) using a micromanipulator. The scanning electron microscope (SEM) micrograph shows an image of the cap layer still attached to the probe (which can be transferred to other photonic structures if desired). What remains on the substrate is a thin diamond membrane, attached to the substrate only at its edges. The thinnest layers produced using this method to date are 210nm, which is achieved using implantation energies of 2.0 and 1.88 MeV helium ions.

An important advantage of this methodology includes the fact that the annealing of the implanted layer takes place with the cap layer in place. As has been shown previously,[33] large pressures are extant in the implanted layer due to the expansion of the damaged layer, primarily due to the change in density from diamond to amorphous and graphitic phases. In essence the cap layer and the underlying substrate act to constrain the sandwich under high pressures during the annealing process and protects the thin membrane from surface oxidation and/or graphitization. We are therefore able to use very high annealing temperatures which would normally lead to graphitization of the desired layer.

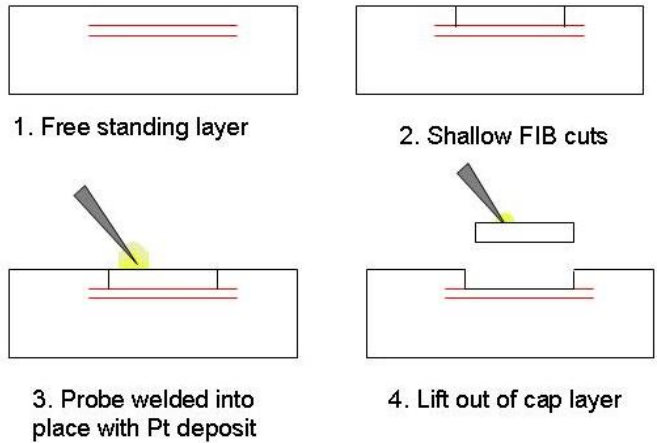
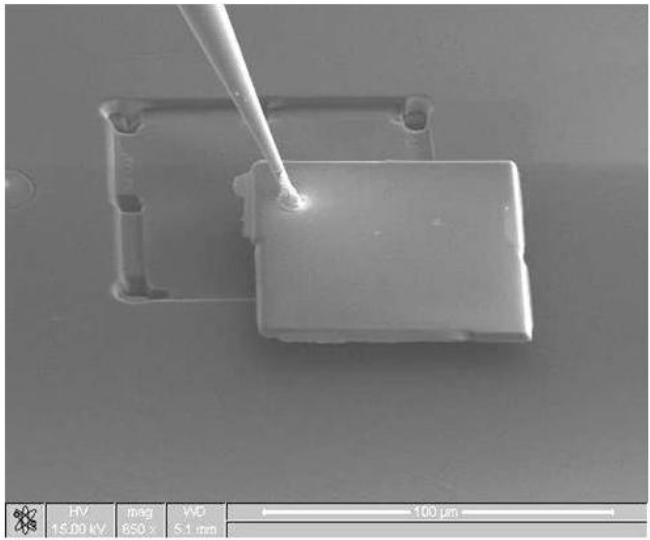

(**Fig. 3**). Schematic of the fabrication procedure: Step 1 shows the sample after annealing and etching with two air layers at depth within the sample. Step 2 shows FIB cuts around ~80% of the edges of the cap layer to enable liftout. Note that the cuts must stop exactly at the shallow layer or the thin membrane below can be lost. Step 3 shows the positioning and attachment of the probe to the cap layer. Step 4, once the final cut is made the cap layer can be removed, exposing the thin membrane beneath for further processing and/or milling. The SEM image shows the freshly exposed layer as the cap is lifted away after being attached to the probe with platinum.

The exposed membranes are then suitable for further processing using FIB to make photonic structures. One such structure is presented in **Fig. 4a**, which shows a micro-ring cavity sculpted from the thin diamond membrane.

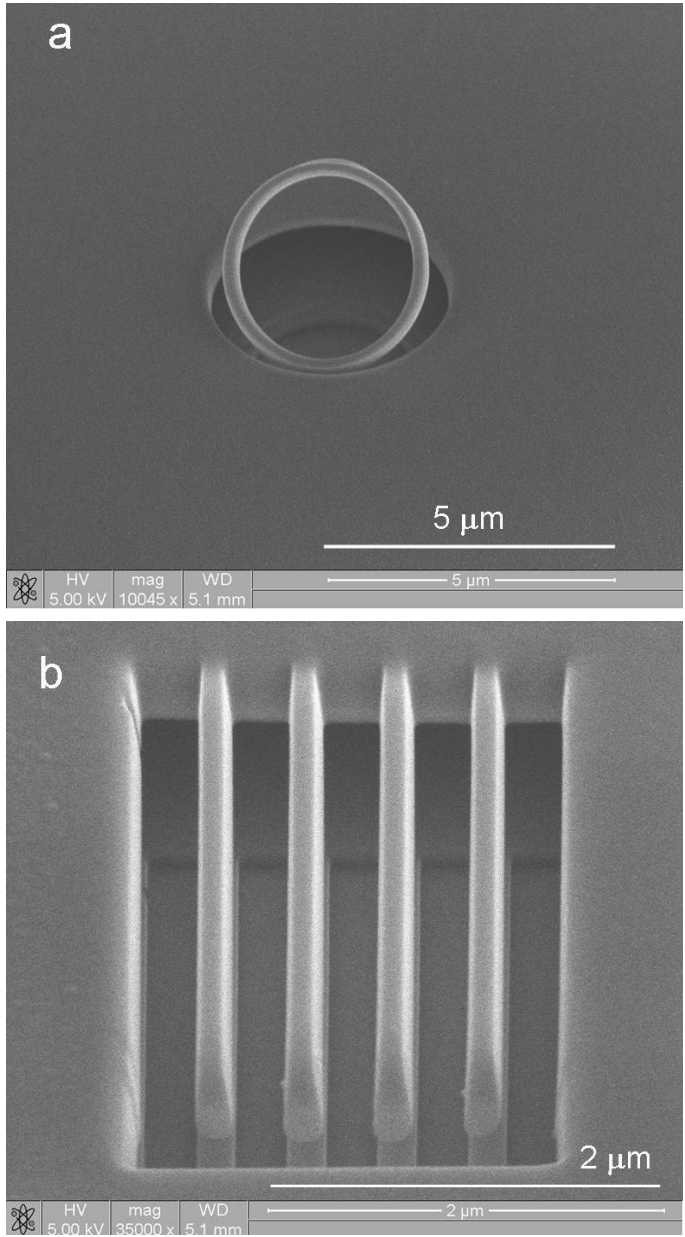

(**Fig. 4**). a) A micron scale ring, which is a prototype for a whispering-gallery mode resonator, fabricated in a 330 nm layer of single crystal diamond. The ring is 3 µm in diameter, with cross section ~280 nm x 330 nm and is attached to the layer by a thin bridge of material. b) Nanocantilevers, 110 nm wide ~4um long in a 330 nm layer. The frequency of vibration for these cantilevers is expected to be of order 1 MHz.

The ring is 3 microns in diameter and only 330 nm in cross section, one of the smallest single crystal diamond rings ever fabricated. Such rings are test structures for future whispering gallery mode micro-resonators. In **Fig. 4b**, cantilevers have been fabricated from the same thin diamond membrane. At this scale there is little or no bending of the free standing cantilevers indicating a low level of residual stress, paving the way for future NEMS devices.

In summary we have shown the fabrication of sub-micron layers of single-crystal diamond suitable for subsequent FIB processing. This method is a significant enabling technology for integrated quantum photonic structures based around colour-centres in diamond. Using our approach we have begun fabrication of a variety of structures including Bragg gratings and whispering gallery mode resonators that are being tested. This method uses a double implant to create two distinct damage layers at depth with in the diamond sample. After annealing, holes are FIB milled to the depth of these layers to enable the etchant solution access to the sacrificial layers. Once etched, the cap can be removed so that further structures can be fabricated in the free standing sub-micron layer. The thinnest layer made to date is 210 nm.

*Experimental*

*MeV Ion Implantation*:

Type Ib diamonds produced by Sumitomo (high-pressure, high-temperature (HPHT) single crystals with a dispersed nitrogen concentration of 10-100 ppm) were implanted with 1.7-2.0, 1.8-2.0, 1.85-2.0 MeV $He^+$ ions combinations on the Blue line of the 5U NEC Pelletron accelerator at the University of Melbourne. The beam was focused to a millimeter spot and scanned with magnetic coils in order to get optimal homogenous ion fluence. The beam current was about 0.5 nA; therefore, the implantation of a 3mm× 3mm region required ~80 min to achieve a fluence of $5.0 \times 10^{16}$ ions.cm$^{-2}$.

*Thermal Annealing*: The annealing was performed in an evacuated chamber, in a vacuum of ~$10^{-5}$ mbar, at ~1260 C for 1 h in a graphite crucible to provide a reducing environment and prevented high-temperature oxidation.

*FIB Milling prior to etching*: Each sample was carbon coated to reduce charging effects, which produced a layer of amorphous carbon on the surface of each sample ~30 nm thick. Milling was performed using a Nova 200 NanoLab from the FEI Company. The FIB provided a 30 keV Ga+ ion beam with a current between 1pA-20 nA depending on milling requirements. Several access holes are milled at this point to enable the etchant solution access to the damaged layers beneath the surface. (The carbon coat is then removed by boiling in a 1:1:1 mix of concentrated Sulphuric ($H_2SO_4$)/Perchloric ($HClO_4$)/Nitric ($HNO_3$) acids for ~5 mins).

*Etching*: The samples were etched using a galvanic etching setup. A saturated Boric acid solution was made and allowed to settle, the etching solution was then made by diluting 1:1 with deionised water daily. A graphite rod (anode) and copper (cathode) electrode were brought close to opposite sides of each sample to increase the effective electric field. The diamond sample to be etched was secured to the base of a Petri dish & just covered with the etching solution. A DC power supply was then used to apply 300V and ~20mA across the sample. Once fully etched the sample was cleaned in deionised water and then boiled in a 1:1:1 mix of concentrated Sulphuric ($H_2SO_4$)/Perchloric ($HClO_4$)/Nitric ($HNO_3$) acids for ~5mins, to remove any remaining graphite residue from the carbon electrode.

*FIB Milling to remove top layer*: Each sample was carbon coated prior to loading into the FIB. At the edge of each implant, wide cuts were then made to remove the top layer. The cuts must be wide enough to allow imaging of the base of the cut as the top layer was thinned away. At a depth of ~2.8 $\mu$m from the surface, the cut begins to approach the thin membrane; real time imaging of the cut area was used to precisely time the completion of the cut. Great care was required on the part of the operator to avoid cutting the submicron layer. Once the layer was free on 80% of the perimeter, a microtip could be brought in contact with the surface. The tip was then welded to the top layer using a platinum depositing system available as a standard add on with the Nova 200 NanoLab. When the tip was secured to the top layer, the final cut freeing the top layer could be made. The top layer can then be removed, exposing the submicron layer below, for further machining.

Figures:

Fig 1:

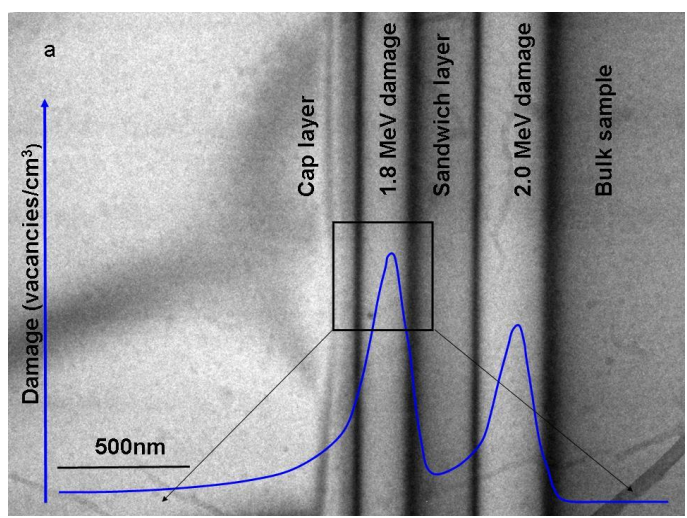

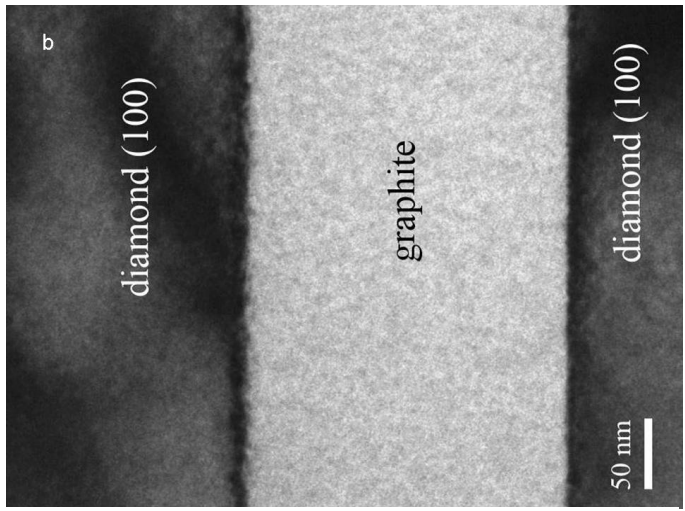

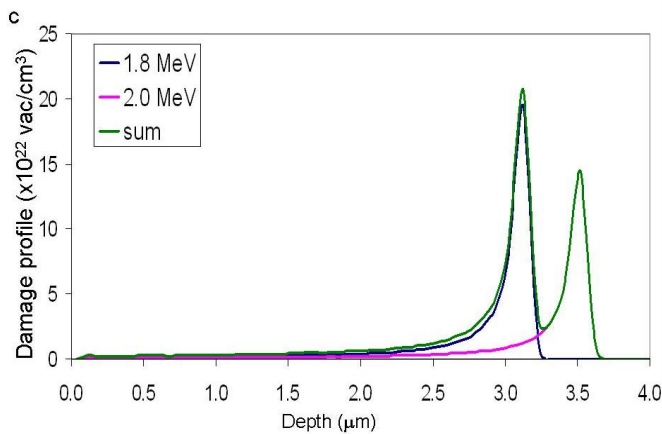

(**Fig. 1a**). Transmission electron microscopy (TEM) image of an unannealed, unetched implant with single crystal diamond and amorphous carbon zones. The sample is implanted from the left in this image. The cap layer and damage zones from the 1.8 and 2.0 MeV helium ions are marked. The 2.0 MeV ions are implanted first with the lower energy ions implanted afterwards. Note the darkened regions between the low damage and high damage regions, TEM indicates this is due to point defects in the diamond lattice.

(**Fig. 1b**). After annealing at 1260C the interface between the damaged and undamaged areas becomes significantly sharper. The highly damaged region converts to graphite and the transition zone between the graphite and diamond appears to be $\leq 10$ nm.

(**Fig. 1c**). SRIM simulation of damage produced from two distinct ion implantations of helium ions into diamond. The 2.0 MeV helium ions, with fluence of $5 \times 10^{16}$ ions/cm$^2$ (pink curve) deposit their energy at greater depth (~3.5 μm) than the 1.8 MeV helium ions, with fluence of $6.8 \times 10^{16}$ ions/cm² (blue curve) at a shallower depth (~3.1 μm). The cumulative damage (green curve) is also shown.

Fig 2:

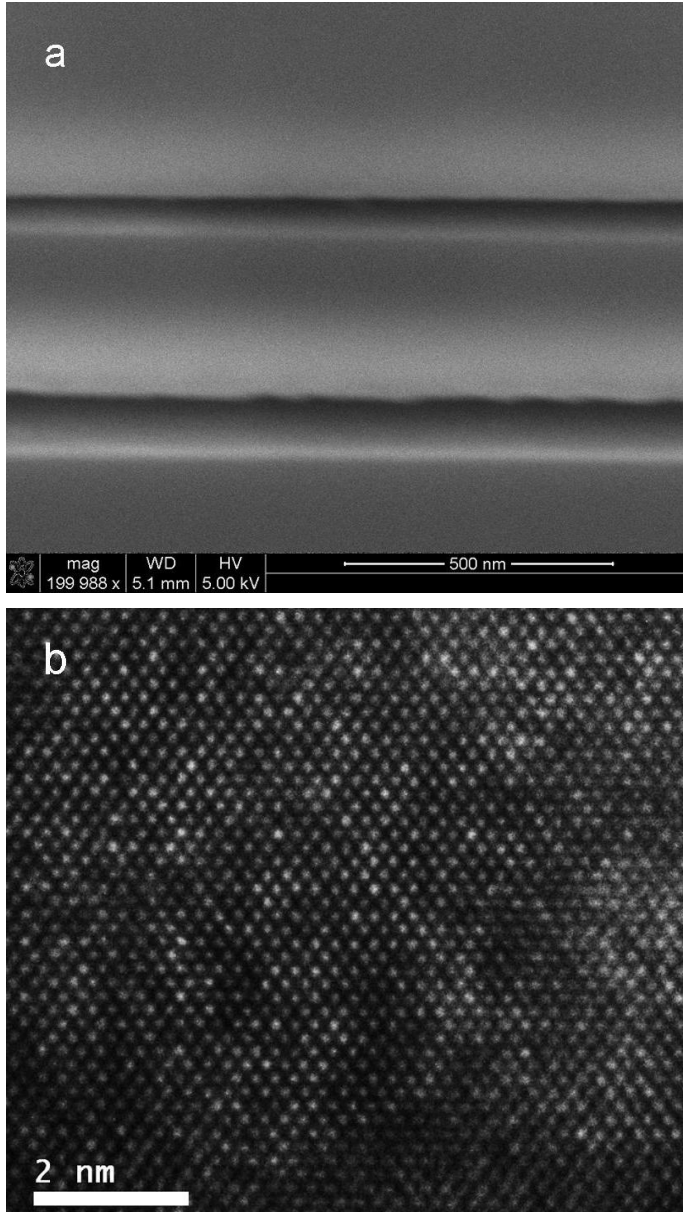

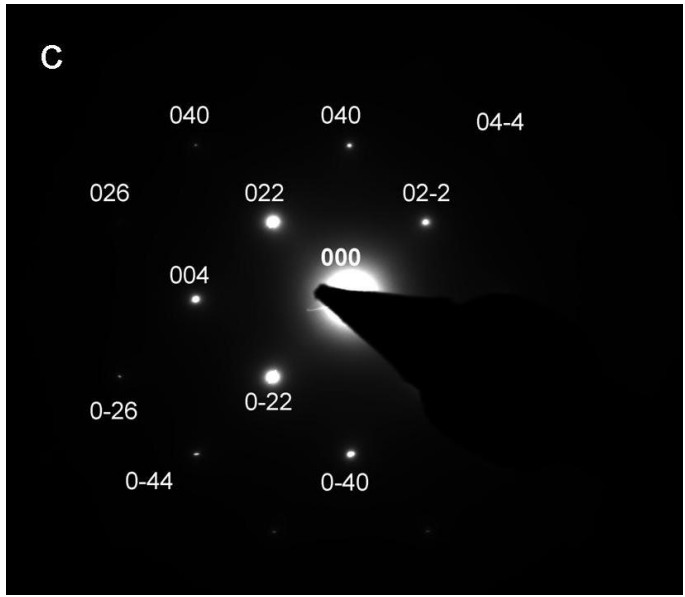

(**Fig. 2**). a) SEM image of 330 nm layer after etching showing ~100 nm air gaps between layers. b) A high resolution TEM (HRTEM) image of diamond layer produced using our method as implanted, before annealing and etching. (c) Diffraction image of diamond membrane after liftout with all spots indexed to single crystal diamond.

Fig 3:

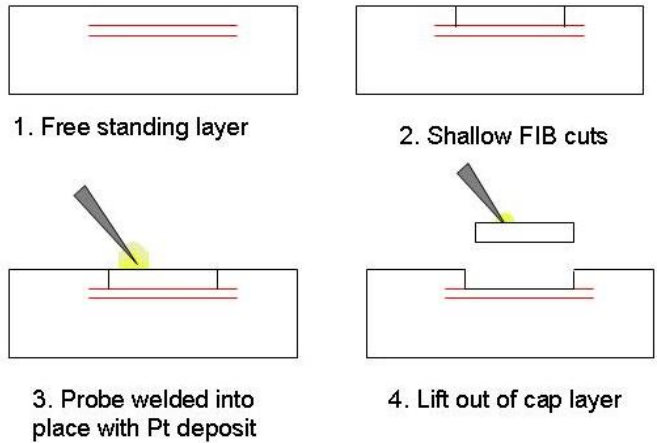

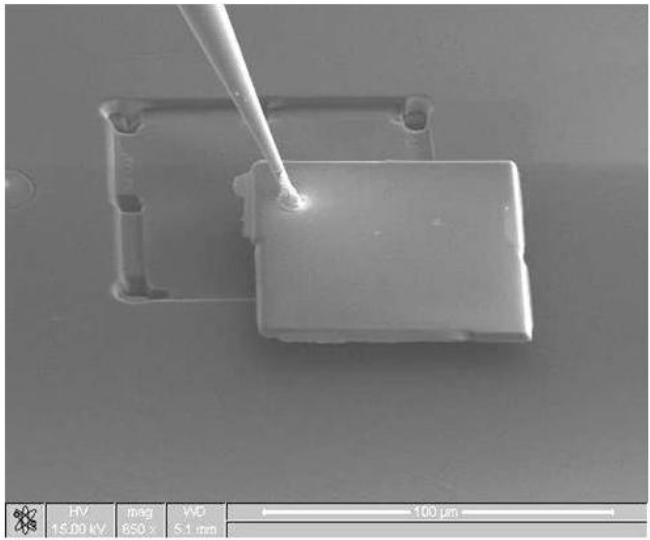

(**Fig. 3**). Schematic of the fabrication procedure: Step 1 shows the sample after annealing and etching with two air layers at depth within the sample. Step 2 shows FIB cuts around ~80% of the edges of the cap layer to enable liftout. Note that the cuts must stop exactly at the shallow layer or the thin membrane below can be lost. Step 3 shows the positioning and attachment of the probe to the cap layer. Step 4, once the final cut is made the cap layer can be removed, exposing the thin membrane beneath for further processing and/or milling. The SEM image shows the freshly exposed layer as the cap is lifted away after being attached to the probe with platinum.

Fig 4:

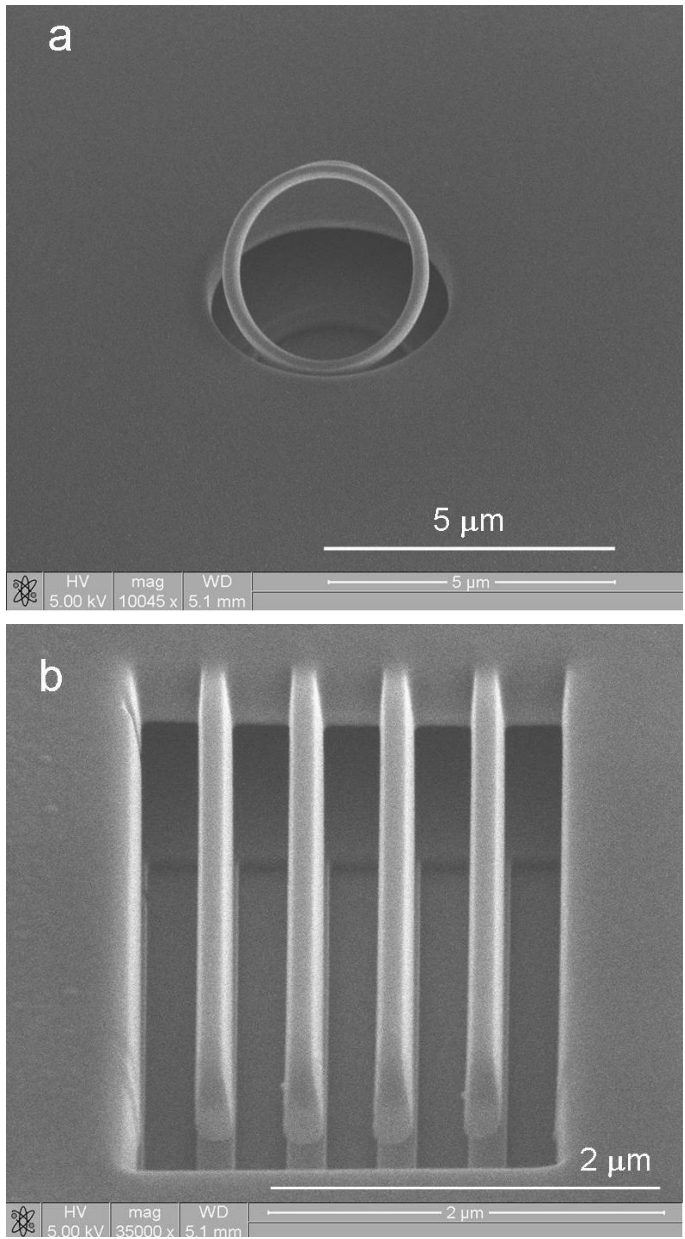

(**Fig. 4**). a) A micron scale ring, which is a prototype for a whispering-gallery mode resonator, fabricated in a 330 nm layer of single crystal diamond. The ring is 3 μm in diameter, with cross section ~280 nm x 330 nm and is attached to the layer by a thin bridge of material. b) Nanocantilevers, 110 nm wide ~4um long in a 330 nm layer. The frequency of vibration for these cantilevers is expected to be of order 1 MHz.

**The table of contents entry** should be fifty to sixty words long, written in the present tense, and refer to the chosen figure.

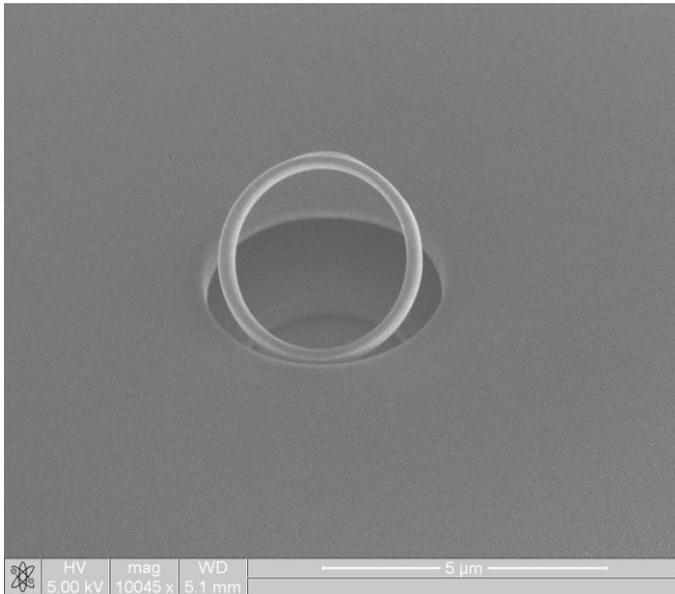

We demonstrate the fabrication of sub-micron layers of single-crystal diamond suitable for subsequent processing as demonstrated by this test ring structure. This method is a significant enabling technology for nanomechanical and photonic structures incorporating colour-centres. The process uses a novel double implant process, annealing and chemical etching to produce membranes of diamond from single-crystal starting material, the thinnest layers achieved to date are 210 nm thick.

Keyword (diamond, sub-micron, photonics, Nano-Electro-Mechanical systems)

B. A. Fairchild Corresponding Author*, P. Olivero, S. Rubanov, A. D. Greentree, F. Waldermann, R. A. Taylor, J. Smith, S. Huntington, B. Gibson, D. N. Jamieson and S. Prawer

Fabrication of ultra thin single crystal diamond membranes